\documentclass[aps,pra,twocolumn,10pt,amsmath,superscriptaddress,amssymb,longbibliography]{revtex4-2}

\usepackage{epsfig}
\usepackage{epstopdf}
\usepackage{dcolumn}
\usepackage{amsbsy}
\usepackage{bm}
\usepackage{color,CJK}
\usepackage{amsmath,amssymb,amsfonts}
\usepackage{appendix}
\usepackage{graphicx}
\usepackage{algorithm}
\usepackage{enumerate}
\usepackage{makeidx}
\usepackage{braket}
\usepackage[usenames,dvipsnames]{xcolor}
\usepackage[driverfallback=dvipdfmx,colorlinks,linkcolor=blue,citecolor=blue,urlcolor=blue]{hyperref}

\begin{document}

\title{Observation of high partial-wave Feshbach resonances in $^{39}$K Bose-Einstein condensates}

\author{Yue Zhang}
\affiliation{State Key Laboratory of Quantum Optics Technologies and Devices, Institute of Opto-electronics, \\Shanxi University, Taiyuan, Shanxi 030006, People's Republic of China }
\affiliation{Collaborative Innovation Center of Extreme Optics, Shanxi University, 
\\Taiyuan, Shanxi 030006, People's Republic of China }

\author{Liangchao Chen}
\email[Corresponding author email: ]{chenlchao87@sxu.edu.cn}
\affiliation{State Key Laboratory of Quantum Optics Technologies and Devices, Institute of Opto-electronics, \\Shanxi University, Taiyuan, Shanxi 030006, People's Republic of China }
\affiliation{Collaborative Innovation Center of Extreme Optics, Shanxi University, 
\\Taiyuan, Shanxi 030006, People's Republic of China }

\author{Zekui Wang}
\affiliation{State Key Laboratory of Quantum Optics Technologies and Devices, Institute of Opto-electronics, \\Shanxi University, Taiyuan, Shanxi 030006, People's Republic of China }
\affiliation{Collaborative Innovation Center of Extreme Optics, Shanxi University, 
\\Taiyuan, Shanxi 030006, People's Republic of China }

\author{Yazhou Wang}
\affiliation{State Key Laboratory of Quantum Optics Technologies and Devices, Institute of Opto-electronics, \\Shanxi University, Taiyuan, Shanxi 030006, People's Republic of China }
\affiliation{Collaborative Innovation Center of Extreme Optics, Shanxi University, 
\\Taiyuan, Shanxi 030006, People's Republic of China }

\author{Pengjun Wang}
\affiliation{State Key Laboratory of Quantum Optics Technologies and Devices, Institute of Opto-electronics, \\Shanxi University, Taiyuan, Shanxi 030006, People's Republic of China }
\affiliation{Collaborative Innovation Center of Extreme Optics, Shanxi University, 
\\Taiyuan, Shanxi 030006, People's Republic of China }

\author{Lianghui Huang}
\affiliation{State Key Laboratory of Quantum Optics Technologies and Devices, Institute of Opto-electronics, \\Shanxi University, Taiyuan, Shanxi 030006, People's Republic of China }
\affiliation{Collaborative Innovation Center of Extreme Optics, Shanxi University, 
\\Taiyuan, Shanxi 030006, People's Republic of China }

\author{Zengming Meng}
\affiliation{State Key Laboratory of Quantum Optics Technologies and Devices, Institute of Opto-electronics, \\Shanxi University, Taiyuan, Shanxi 030006, People's Republic of China }
\affiliation{Collaborative Innovation Center of Extreme Optics, Shanxi University, 
\\Taiyuan, Shanxi 030006, People's Republic of China }

\author{Ran Qi}
\affiliation{Department of Physics, Renmin University of China, Beijing 100872, People’s Republic of China }

\author{Jing Zhang}
\email[Corresponding author email: ]{jzhang74@sxu.edu.cn}
\affiliation{State Key Laboratory of Quantum Optics Technologies and Devices, Institute of Opto-electronics, \\Shanxi University, Taiyuan, Shanxi 030006, People's Republic of China }
\affiliation{Collaborative Innovation Center of Extreme Optics, Shanxi University, 
\\Taiyuan, Shanxi 030006, People's Republic of China }

%\date{\today }

\begin{abstract}
We report the new observation of several high partial-wave (HPW) magnetic Feshbach resonances (FRs) in $^{39}$K atoms of the hyperfine substate $\left|F=1,m_{F}=-1\right\rangle$. These resonances locate at the region between two broad $s$-wave FRs from 32.6 G to 162.8 G, in which Bose-Einstein condensates (BECs) can be produced with tunable positive scattering length obtained by magnetic FRs. These HPW FRs are induced by the dipolar spin-spin interaction with s-wave in the open channel and HPW in the closed channel. Therefore, these HPW FRs have distinct characteristics in temperature dependence and loss line shape from that induced by spin-exchange interaction with HPWs in both open and closed channels. Among these resonances, one $d$-wave and two $g$-wave FRs are confirmed by the multichannel quantum-defect theory (MQDT) calculation. The HPW FRs have significant applications in many-body physics dominated by HPW pairing.
\end{abstract}
\maketitle

Feshbach resonance is a powerful tool to flexibly tune the atomic interaction in researches of ultracold atomic gases. Utilizing FRs, ultracold molecules~\cite{nature.426.537, science.301.1510, nature2021.592}, many-body physics~\cite{rmp.80.885, rmp.89.035006} (also including quantum droplets~\cite{science.359.301, prl.120.235301}, matter-wave soliton~\cite{science.356.422}, and BEC-BCS crossover~\cite{prl.92.120401, pr412.1}) , and out-of-equilibrium dynamics~\cite{np.13.704, prl.118.230403, np.8.213, arocmp.6.201} have been studied. We usually call a specific potential energy curve between two colliding atoms a scattering channel, and FRs result from the coupling between the open and closed channel. By adjusting an external magnetic or optical field, the energy of the scattering atoms in the open channel can be brought close to that of the molecular bound state in the closed channel. The two channels can be coupled under different interaction schemes, and then the scattering properties of free atoms in the open channel are significantly changed by the bound states in the closed channel~\cite{rmp2010.82}. The scattering wave function of two colliding atoms can be expanded into different partial-waves, each with a specific orbital angular momentum $l$ and a projection $m_{l}$ of $l$. When $l = 0,1,2,3...$, the corresponding partial-wave is called $s, p, d, f... $ wave~\cite{pra.76.042514}. If both the open and closed channels are $s$-wave, the resonance is called $s$-wave FR, while it is called HPW FR with a non-zero partial-wave in the open channel or the closed channel.

The interaction scheme between atoms determines the coupling strength and selection rules of the various partial-waves in the open and closed channels. There are two main types of interaction: the spin-exchange interaction and dipolar spin-spin interaction. Both of them can induce $s$-wave or HPW FRs, and we will mainly focus on the HPW FRs in this paper.

\textbf{1) }The common stronger coupling in FR arises from the spin-exchange interaction between two atoms. Such interaction is isotropic, and the total orbital angular momentum as well as its projection of two atoms are conserved in collisions. Then the selection rule over the coupling between partial waves of the open and closed channels is $\Delta l = 0$, $\Delta m_{l} = 0$, $\Delta m_{f} = 0$.  Here $m_{f}$ is the projection of the total spin angular momentum $f$. Therefore, for HPW FRs induced by spin-exchange interaction, the scattering waves in both the open and closed channels are HPWs, and every $m_{l}$ component in the open channel can find an equal $m_{l}$ component in the closed channel to form an effective coupling. Since the different $\lvert m_{l} \rvert $ can split the energy levels of molecular bound states in the closed channel, such HPW FRs will have a characteristic $l+1$ splitting structure. As a good example of such phenomena, a triplet structure of $d$-wave FR can be resolved in the atomic loss spectrum of ultracold $^{85}$Rb and $^{87}$Rb mixture~\cite{prl.119.203402}. In addition, the HPWs in the open channel will add centrifugal potential energy to the atomic interaction, resulting in the strong temperature dependence of FRs~\cite{pra.69.042712, pra.98.022704}. Besides, the atomic loss spectrum tends to display asymmetric line shapes because of the HPWs in open channels~\cite{pra.99.022701, prl.119.203402, pra.98.022704,np2019.15}.

%%%%%%%%%%%%%***Fig. 1***%%%%%%%%%%%%%%%%%%
\begin{figure*}[htbp]
\centerline{\includegraphics[scale=1.0]{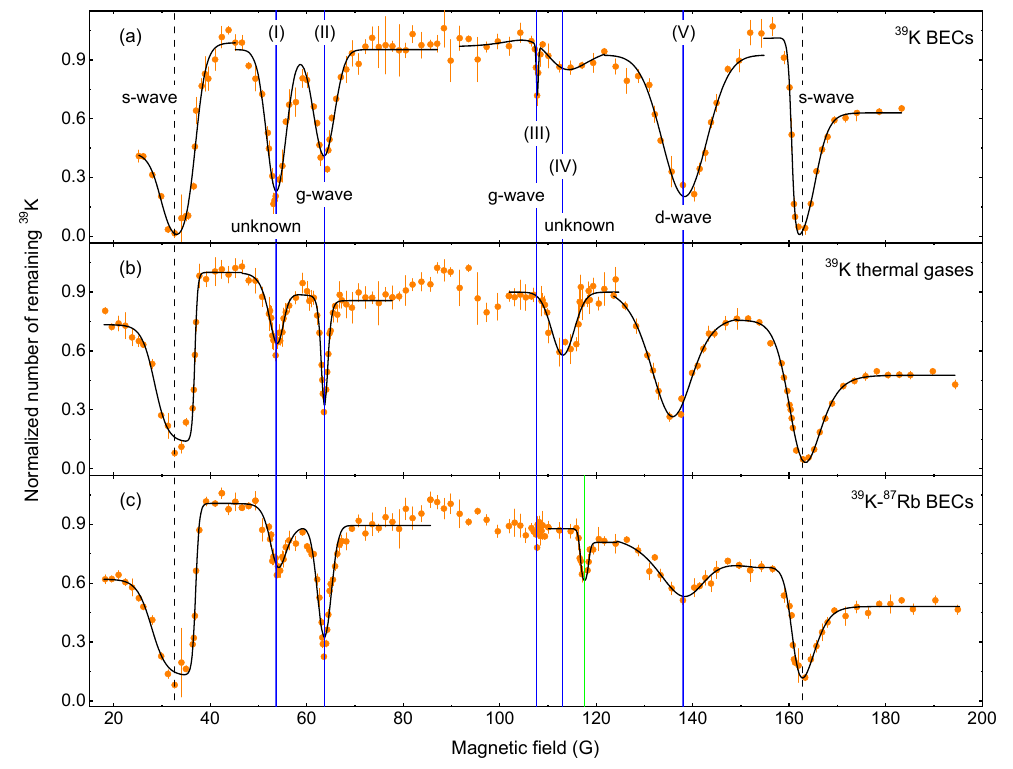}}
\caption{(Color online). Atomic loss spectra of $^{39}$K in $\left|1,-1\right\rangle$ state. Five newly observed FRs can be seen in the spectra. (a) Spectrum for $^{39}$K BECs. (b) Spectrum for thermal $^{39}$K atomic gases. (c) Spectrum for $^{39}$K-$^{87}$Rb mixture BECs. These FRs are labeled as (I), (II), (III), (IV), and (V) respectively. The black dash lines indicate the previously reported $s$-wave FRs, the blue solid lines indicate the resonance centers of our newly observed $^{39}$K FRs, the green solid line indicates the $^{39}$K-$^{87}$Rb interspecies FR. Each data point for all the measurements in this paper is repeated 3 times, the error bars represent the standard error. The black solid lines show the fits of the experimental data, the symmetric loss dips are fitted with the Gaussian function, and the asymmetric dips are fitted with the double sigmoidal function. }
\label{fig.1}
\end{figure*}
%%%%%%%%%%%%%****Fig. 1***%%%%%%%%%%%%%%%%%%

\textbf{2) }Another important interaction is the dipolar spin-spin interaction. The spin-spin interaction is usually weaker and strongly anisotropic so that the orbital angular momentum $l$ and the spin angular momentum $f$ can be mutually converted in the collision, while the total angular momentum projection $m_{l} + m_{f}$ is conserved. So this interaction gives a selection rule $\lvert \Delta l \rvert = 0, 2, 4$ and $\Delta m_{l} + \Delta m_{f} = 0$~\cite{prl.119.203402, pra.99.022701, pra.76.042514, pra.83.042704, pra.69.042712}. In ultracold atomic gases, the dipolar spin-spin interaction can induce HPW FRs with $s$-wave in the open channel and HPW (usually a $d$-wave or $g$-wave) in the closed channel. Then the $s$-wave in the open channel (in which $m_{f}=m_{fo}$, $m_{l}=m_{lo} = 0$) can only be coupled with one component of the HPW in the closed channel (in which $m_{f}=m_{fc}$, $m_{l} =m_{fo}-m_{fc}$) according to the selection rules when the substates of the same hyperfine state are largely nondegenerate in magnetic fields (so $m_{fo}$ and $m_{fc}$ are fixed values for a specific FR ). So, this kind of HPW FRs has no multiple splitting structures. Such type of HPW FRs has been observed in many atomic gases~\cite{pra.83.042704, prl.89.283202, pra.87.033611, pra.87.032517}. Since there is no centrifugal potential barrier in the open channel to prevent atoms from approaching each other, the line shapes of these HPW FRs have very different temperature dependence and are usually symmetric.

HPW FRs have been observed in atomic gases of many alkali metal isotopes, such as $^{6}$Li~\cite{pra.70.030702, pra.71.045601}, $^{23}$Na~\cite{pra.83.042704}, $^{40}$K~\cite{prl.95.230401, prl.98.200403, np2016.12}, $^{41}$K~\cite{pra.98.022704, np2019.15}, $^{85}$Rb-$^{87}$Rb~\cite{pra.94.062702, prl.119.203402}, $^{133}$Cs~\cite{pra.76.042514, prl.100.083002, nature2021.592}. In addition, it has also been demonstrated in degenerate Fermi gas of $^{40}$K that, the synthetic spin-orbit coupling can induce $p$-wave pairing through $s$-wave FRs~\cite{np2014.10}. These HPW FRs will help greatly to accurately calibrate the potential energy curve between atoms~\cite{pra.77.052705, pra.83.042704, pra.98.062712, X1912.01264}, and find novel Fermion~\cite{prl.95.070404, prl.94.230403, prl.94.090402, prl.99.210402} or Boson~\cite{pra.97.043626, nature2021.592, X2101.05431} superfluid states with nonzero orbital angular momentum in ultracold atomic gases. The HPW pairing plays a decisive role in some intriguing superfluids~\cite{rmp1997.69} and high-temperature superconductors~\cite{nature1994.372, prl.85.182}, and the related research in ultracold atoms can deepen our understanding of these problems.

%%%%%%%%%%%%%***FIG. 2***%%%%%%%%%%%%%%%%%%
\begin{figure}[htbp]
\centerline{\includegraphics[scale=1.0]{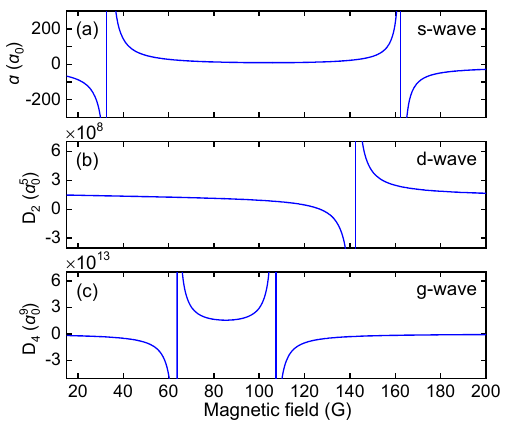}}
\caption{(Color online). Scattering lengths (hyper-volumes) of $^{39}$K in $l=0,2,4$ channels calculated within MQDT.}
\vspace{-1 em}
\label{fig.2}
\end{figure}
%%%%%%%%%%%%%***FIG. 2***%%%%%%%%%%%%%%%%%%

FRs of $^{39}$K in various spin states have been specifically studied in the references~\cite{prl.111.053202, pra.81.032702, njp2007.9, prr.2.013366, prr.5.013174}, where most of them are the $s$-wave resonances. A $d$-wave FR in 394 G has been observed for thermal $^{39}$K gases in $\left|1,-1\right\rangle$ state~\cite{pra.99.022701}, however the negative background scattering length makes it difficult to produce BECs in this region. This FR has HPWs in both the open and closed channels, and the triplet splitting structure as well as asymmetric line shape can be seen in the atomic loss spectrum. $^{39}$K atoms have wide FRs in the relatively lower magnetic field strength, making it a suitable system for investigating many-body physics. Utilizing the tunable atomic interaction in $^{39}$K BECs, fascinating researches like quantum droplets~\cite{science.359.301, prl.120.235301}, artificial gauge field~\cite{nature608.293}, Anderson localization~\cite{np11.554, nature453.895}, Efimov states~\cite{prl.125.243401, prl.123.233402, prr.5.013174}, mobile polarons~\cite{prl.117.055302} have been studied.

In this paper, we report our experimental observation of new HPW FRs of $^{39}$K BECs in $\left|1,-1\right\rangle$ state, which have $s$-wave in open channel and HPW in closed channel induced by the dipolar spin-spin interaction.  Our experimental setup producing the $^{39}$K-$^{87}$Rb dual-species BECs has been previously described elsewhere~\cite{cpb2021.30}, and we only give a brief introduction here. A dark $^{87}$Rb three-dimensional magneto-optical trap (3D MOT) and a bright $^{39}$K 3D MOT collect atoms transferred from a 2D MOT. After the molasses (gray molasses for $^{39}$K) and optical pumping procedure, the atomic mixture is loaded into a blue-detuning optically plugged quadrupole magnetic trap, in which $^{87}$Rb atoms are evaporatively cooled by microwave radiation (frequency sweeping is truncated at 6.8370 GHz), and $^{39}$K atoms are sympathetically cooled with $^{87}$Rb atoms. Atoms in the optically plugged quadrupole trap are then transferred adiabatically into a crossed optical dipole trap formed by two 1064 nm laser beams. By this stage, both $^{87}$Rb and $^{39}$K atoms in the dipole trap are in the $\left|2,2\right\rangle$ state. Then the $^{87}$Rb atoms are transferred into $\left|1,1\right\rangle$ state by sweeping the frequency of the microwave radiation around 6.8412 GHz and successively the $^{39}$K atoms are transferred into $\left|1,1\right\rangle$ state by sweeping another microwave radiation around 468.58 MHz. In experiments, it is necessary to transfer the $^{87}$Rb into $\left|1,1\right\rangle$ state before $^{39}$K atoms to avoid the spin-exchange heating. The transfer efficiency for both atoms is about $95\%$, and the remanent few atoms in $\left|2,2\right\rangle$ state are blown away by a resonant flash light.

%%%%%%%%%%%%%***Table I***%%%%%%%%%%%%%%%%%%
\begin{table}[tbp]
\vspace{-1 em}
\renewcommand{\arraystretch}{1.2}
\setlength{\belowcaptionskip}{1.5mm} %
\centering \caption{Physical quantities of the atomic gases prepared for the measurement in Fig.~\ref{fig.1}. These gases have an atomic number $N$, a temperature $T$, and the optical dipole trap has a geometric mean trap frequency of $f_{\mathrm{d}}$.}
\setlength{\tabcolsep}{1.8mm} 
\begin{tabular}{c|ccc}
\hline
\hline 
Samples & $N$ & $T$ (nK) & $f_{\mathrm{d}}$ (Hz) \tabularnewline
% \multicolumn{1}{c|}{Samples} & \multicolumn{3}{c}{atomic number & temperarure (nK) & trap frequency (HZ) }
\hline 
$^{39}$K BECs & $3.0(2) \times 10^{5}$ & 30(5) & 70(1) \tabularnewline
 $^{39}$K thermal gases & $2.5(3) \times 10^{5}$ & 160(23) & 70(1) \tabularnewline
\hline 
$^{39}$K in $^{39}$K-$^{87}$Rb BECs & $2.9(2) \times 10^{5}$  & 40(4) & 70(1) \tabularnewline
$^{87}$Rb in $^{39}$K-$^{87}$Rb BECs & $2.6(3)\times10^{5}$ &30(2) &  55(1) \tabularnewline
\hline 
\hline 
\end{tabular}
\vspace{0 em}
\label{tab.I}
\end{table}
%%%%%%%%%%%%%***Table I***%%%%%%%%%%%%%%%%%%

By gradually decreasing the intensity of the dipole trap, the atoms are evaporatively cooled further. Since the $^{87}$Rb atoms are heavier than the $^{39}$K atoms, they will escape faster from the dipole trap during the evaporative cooling. When the temperature of $^{87}$Rb reaches close to the critical temperature of Bose-Einstein condensation, a radio frequency field is applied to transfer both atoms from $\left|1,1\right\rangle$ to $\left|1,-1\right\rangle$ state simultaneously in a bias magnetic field of 10 G. The $^{39}$K atoms have a negative $s$-wave background scattering length, which will prevent the $^{39}$K atoms from Bose-Einstein condensation. Therefore we set the magnetic field to a suitable region, where two broad $s$-wave FRs lead to a positive scattering length, such that stable BECs can be produced. Then, the evaporative cooling toward BECs is performed further, during which the magnetic field is optimized to be 120.18 G near a $^{39}$K-$^{87}$Rb interspecies FR at 117.59 G. The larger interspecies scattering length leads to a higher sympathetic cooling efficiency, which helps to increase the atomic number in the finally obtained $^{39}$K BECs. The gravitational sag will reduce the overlap of $^{87}$Rb and $^{39}$K atomic clouds in the weaker dipole trap. Therefore, it takes a longer time for $^{39}$K atoms to condense in the last step of evaporative cooling. Finally, after a 4 s evaporation, we obtain dual-species BECs of $^{39}$K and $^{87}$Rb in $\left|1,-1\right\rangle$ state. We have checked that no other spin states of the two species have been left in the mixtures. Especially for $^{39}$K atoms, only $\left|1,-1\right\rangle$ state has a positive scattering length at 120.18 G and can exist in BECs for a long time. To obtain a single $^{39}$K BECs, we can blow $^{87}$Rb atoms away by resonant flash light, or leak all the $^{87}$Rb atoms out of the dipole trap by ramping down the dipole trap depth further.

%%%%%%%%%%%%%***FIG. 3***%%%%%%%%%%%%%%%%%%
\begin{figure*}[htbp]
\centerline{\includegraphics[scale=1.0]{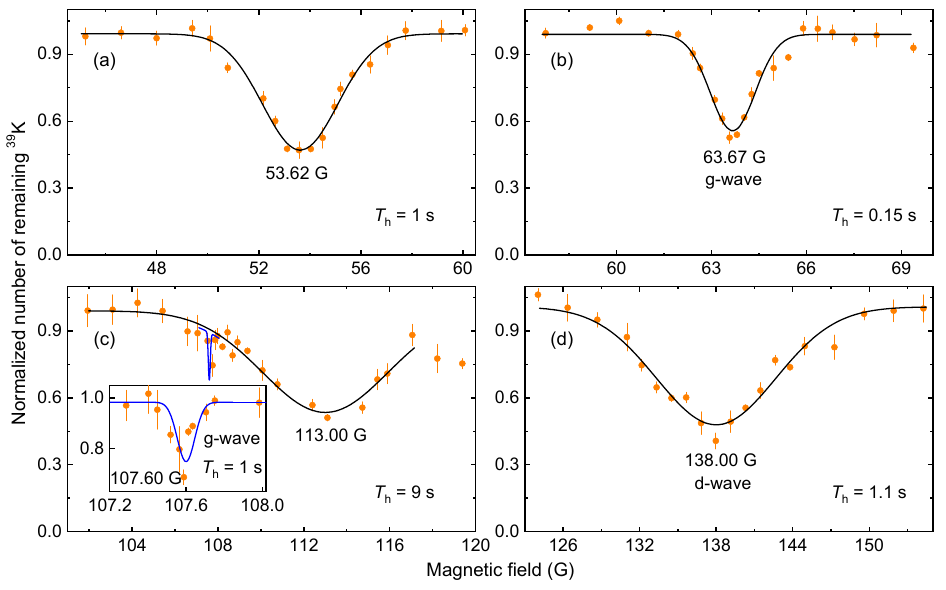}}
\caption{(Color online). Atomic loss spectra around the five new FRs in $^{39}$K BECs. The holding times ($T_{\mathrm{h}}$) in the corresponding magnetic field are properly set for each spectrum to have comparable loss dips between them. The measured resonance centers, loss dip widths, and the holding times for each spectrum are summarized in Table~\ref{tab.II}.}
\label{fig.3} 
\end{figure*} 
%%%%%%%%%%%%%***FIG. 3***%%%%%%%%%%%%%%%%%%

After producing the single $^{39}$K BECs, we scan the magnetic field from 20 G to 200 G to measure the atomic loss accordingly. The magnetic field has been calibrated by radio frequency spectroscopy of $^{87}$Rb atomic transition from $\left|1,1\right\rangle$ to $\left|1,0\right\rangle$ state. The magnetic field has a short-term fluctuation of about 10 mG and a long-term drift of tens of mG.  We ramp the magnetic field from 120.18 G to the target points in 30 ms, and it will settle down in less than 20 ms. After holding the magnetic field for 2 s, we switch off the dipole trap to set a free expansion of the atomic gases for 5 ms while keeping the magnetic field unchanged. During this time, the energy of the BECs in the target magnetic field is released to kinetic energy, and the density of the atomic gases is decreased such that the negative background scattering length close to the zero magnetic field will not induce additional atomic loss when we switch off the magnetic field before absorption imaging. Finally, after switching off the magnetic field and waiting for another 20 ms time of flight, we take the absorption imaging to measure the number of residual atoms.

Fig.~\ref{fig.1}(a) shows the atomic loss spectrum of $^{39}$K BECs. The quantities of the gases have been summarized in Table~\ref{tab.I}. Besides the two already known $s$-wave FRs in 32.6 G and 162.8 G, there are another five resonant loss regions. Based on the analysis of MQDT, resonance (V) is identified as $d$-wave FR, resonance (II) and (III) are $g$-wave FR, and resonance (I) and (IV) are unclear now.

The MQDT method adopted in this paper was described in detail in~\cite{pra.98.022704,np2019.15} and the supplementary material therein. In MQDT, quantum defect functions $K_{S,T}^{c}(\epsilon,l)$ are used to connect the long-range asymptotic behavior of scattering atoms with the complex short-range interaction. We adopt the same form of $K_{S,T}^{c}(\epsilon,l)$ as used in~\cite{pra.98.022704}, in which $\beta_{S,T}$ parameters are introduced to include the angular momentum dependence of $K_{S,T}^{c}(\epsilon,l)$:
\begin{align}
K_{S,T}^{c}(\epsilon,l)=K_{S,T}^{c}(0,0)+\beta_{S,T}l(l+1)
\end{align}
 For the calculation of FRs in $^{39}$K, we take $K_{S}^{c}(0,0)=1.7631,~K_{T}^{c}(0,0)=-0.1999$ and $\beta_S=-0.0082,~\beta_T=-0.0010$ to get a best agreement with experimental data. With this set of parameters, we can reproduce the resonance position of all known FRs reported in earlier references and the HPW parameters $\beta_{S,T}$ are necessary to reproduce the new FRs we found.

%%%%%%%%%%%%%***Table II***%%%%%%%%%%%%%%%%%%
\begin{table*}[htbp]
\vspace{-1 em}
\renewcommand{\arraystretch}{1.2}
\setlength{\belowcaptionskip}{1.8mm} %

\centering \caption{Observed FRs in our experiments. $B_{\mathrm{expt}}$ is the resonance center extracted from the loss spectrum; $B_{\mathrm{th}}$ and $\Delta_{\mathrm{th}}$  are resonance center and width obtained from the MQDT calculation; $a_{l\mathrm{bg}}$ is the characteristic length representing the background hyper-volume $D_{l\mathrm{bg}}$ (see the text), and $a_{0}$ is the Bohr radius. Widths $\Delta_{\mathrm{loss}}$ of the loss spectra of FRs in Fig.~\ref{fig.3} and the carefully chosen holding times $T_{\mathrm{h}}$ are also listed in the table. All the HPW FRs reported in this paper have an $s$-wave in the open channel.}

\setlength{\tabcolsep}{2.0mm} 
\begin{tabular}{c|cc|ccc|ccc}
\hline 
\hline 
\multicolumn{1}{c|}{FRs in Fig.~\ref{fig.1} }&\multicolumn{2}{c|}{Bound states} & \multicolumn{3}{c|}{MQDT calculation}&\multicolumn{3}{c}{fitting in Fig.~\ref{fig.3}}\tabularnewline
 (partial-wave)&$l$ & $m_{l}$& $B_{\mathrm{th}}$ (G) & $\Delta_{\mathrm{th}}$ (G) & $a_{l\mathrm{bg}}$ ($a_{0}$)  &  $B_{\mathrm{expt}}$ (G) & $\Delta_{\mathrm{loss}}$ (G) & $T_{\mathrm{h}}$ (s)\tabularnewline
\hline 
($s$-wave)~\cite{njp2007.9} & 0 & 0 & 33.60 & 55 & -19& 32.6(1.5)  &- &- \tabularnewline
($s$-wave)~\cite{njp2007.9} & 0 & 0 & 162.30 & -37 & -19& 162.8(9)  &- &- \tabularnewline
($s$-wave)~\cite{njp2007.9} & 0 & 0 & 560.7 & -56 & -29& 562.2(1.5)  &- &- \tabularnewline
\hline 
I (unknown) & -& - &- &- &- & 53.62(7) & 3.47(20) & 1\tabularnewline
II ($g$-wave) & 4 & 0 & 63.40 &-43&25.1& 63.67(9)  & 1.67(20) & 0.15 \tabularnewline
III ($g$-wave) & 4 & 0 & 106.82 &42&25.1& 107.60(2)  & 0.11(4) & 1\tabularnewline
IV (unknown) & -& -&- & -&- & 113.00(13) & 6.88(38) & 9\tabularnewline
IV ($d$-wave) & 2 & 0 & 141.84 &-16&42.1 & 138.00(19) & 11.00(1.02) & 1.1 \tabularnewline
\hline 
\hline 
\end{tabular}
\vspace{0 em}
\label{tab.II}
\end{table*}
%%%%%%%%%%%%%***Table II***%%%%%%%%%%%%%%%%%%

For an FR with arbitrary partial-wave $l$, the scattering amplitude is characterized by a generalized scattering hyper-volume $D_{l}(B)$. When magnetic field $B$ is close to the resonance position $B_0$, $D_{l}(B)$ can be cast into the following simple form
\begin{align}
D_{l}(B)=D_{l\mathrm{bg}}\left(1-\frac{\Delta}{B-B_0}\right),
\end{align}
in which one can define the scattering width $\Delta$ and the background hyper-volume $D_{l\mathrm{bg}}$. For $s$-wave FRs ($l=0$), $D_0(B)$ reduces to the scattering length $a(B)$ while $D_{0\mathrm{bg}}$ is simply the background scattering length $a_{\mathrm{bg}}$. For arbitrary $l$, $D_{l}(B)$ and $D_{l\mathrm{bg}}$ have the dimension of $[\mathrm{Length}]^{2l+1}$ and we define $a_{l\mathrm{bg}}=\mathrm{sign}(D_{l\mathrm{bg}})D_{l\mathrm{bg}}^{1/(2l+1)}$ as a characteristic length scale representing $D_{l\mathrm{bg}}$. The $D_{l}(B)$ for $l=0,2,4$ can be calculated within MQDT and the results are shown in Fig.~\ref{fig.2}. The quantum numbers describing the molecular bound states of each FR observed in the current experiment are summarized in Table~\ref{tab.II}. The measured resonance centers and corresponding theoretical results obtained from MQDT are also compared in the table.

We also prepare a thermal sample of the $^{39}$K atoms by adjusting the sympathetic cooling in the dipole trap. Then we repeat the atomic loss measurement, which gives the result shown in Fig.~\ref{fig.1}(b). It can be seen that the response of those five resonances to temperature is quite different. The dips of FR (I, III, V) become shallower, the dip of FR (IV) becomes deeper, and the dip of $g$-wave FR (II) remains almost unchanged. The atoms collide in s-waves in the open channels of these FRs, so they do not need enough initial kinetic energy to cross the centrifugal barrier to couple with the molecular states. Besides, for atomic gases in the same dipole trap, an increase in temperature leads to a decrease in the atomic number density and a decrease in the collision rate of atoms. Therefore, it seems reasonable that FRs (I, III, V) become shallower with increasing temperature. It is also understandable that FR (II) is almost unchanged. It can be seen from Fig.~\ref{fig.3} that FR (II) has a relatively stronger coupling strength (minimum holding time), so FR (II) can still cause strong atomic loss in a long enough holding time despite the decrease in atomic number density. The response of FR (II) is consistent with that of the two strong s-wave FRs (32.60 G, 162.80 G). However, FR (IV) responds to the changed temperature in a very different way. FR (IV) may be a different type of FR from the others. We expect that the rich behaviors of these FRs will allow examining the universal law of three-body collisions~\cite{prl.111.053202, prl.125.243401, prl.123.233402, prr.5.013174} near the HPW FRs. In addition, it should be noted that for FRs with HPW in both open and closed channels, the multiple splitting structure on the loss spectrum will change with temperature~\cite{prl.119.203402, np2019.15}, which is very different from the HPW FRs we observed.

In Fig.~\ref{fig.1}(c), we show the loss spectrum of $^{39}$K-$^{87}$Rb mixture BECs, in which the numbers of remaining $^{39}$K atoms in different magnetic field strengths are recorded. From this figure, we can see that the five new FRs also have different responses to the mixing of $^{87}$Rb atoms. Note that,  in the magnetic field range of our experiments, only one interspecies FR between $^{39}$K and $^{87}$Rb atoms in the $\left|1,-1\right\rangle$ state has been reported in the references and was observed in our experiment, which has a resonance center at 117.56 G, a width of 1.21 G, and a background scattering length of 28.5 $a_{0}$~\cite{pra.92.053602}. The interspecies scattering length near the five new FRs is small but comparable with the intraspecies scattering length of $^{39}$K atoms. Furthermore, the $^{87}$Rb BECs have a larger scattering length of 100 $a_{0}$ and a large atomic number. Thus, it is reasonable to assume that the $^{87}$Rb atoms act as buffer gases, reducing the collision rate of two $^{39}$K atoms. And in the region out of the interspecies FR, the collision loss between $^{87}$Rb and $^{39}$K atoms is much smaller than that of two $^{39}$K atoms at FRs. As a result, we can see that the loss dips of FRs (I, III, IV, V) become shallower when mixed with $^{87}$Rb atoms. The dip of FR (II) remains almost unchanged still because it is relatively stronger than others. Comparing the loss dips of FR (IV) in Fig.~\ref{fig.1}(a, b, c), we can infer that FR (IV) is very sensitive to temperature (more specifically, the collision energy of free atoms).

%%%%%%%%%%%%%***FIG. 4***%%%%%%%%%%%%%%%%%%
\begin{figure*}[!t]
\centerline{\includegraphics[scale=1.0]{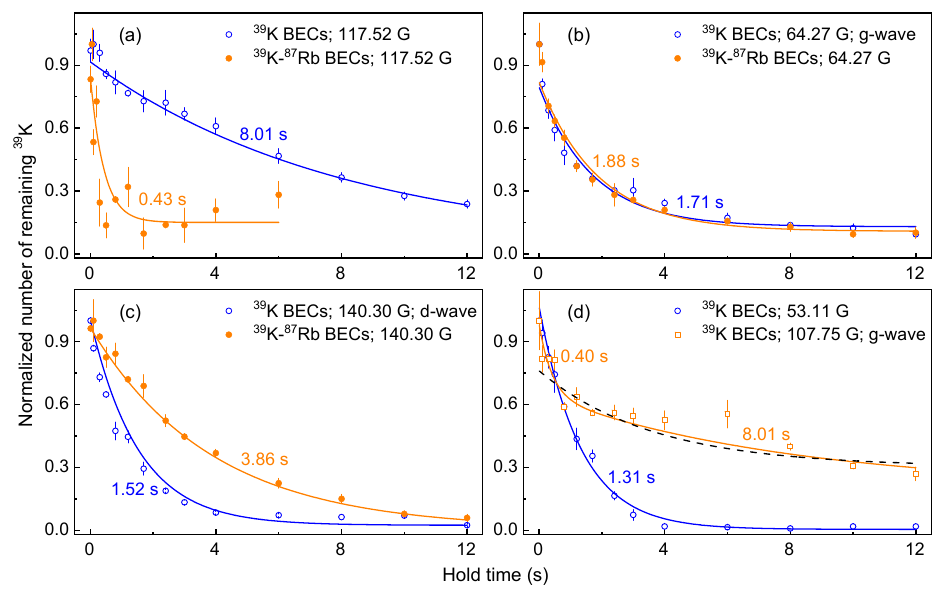}}
\caption{(Color online). Atomic loss process of $^{39}$K BECs and $^{39}$K-$^{87}$Rb mixture BECs around FRs. (a) Magnetic field strength is set to $B=117.52$ G, close to a $^{39}$K-$^{87}$Rb interspecies FR. (b) $B=64.27$ G, close to a $g$-wave FR. (c) $B=140.30$ G, close to a $d$-wave FR. (d) The open circles correspond to $B=53.11$ G, and the open squares correspond to $B=107.75$ G, close to a $g$-wave FR. In all the above figures, the open circles and squares represent single $^{39}$K BECs, and the solid circles represent $^{39}$K-$^{87}$Rb mixture BECs. Most experimental data are fitted well with a single exponential function, as shown by the solid lines, and the fitted atomic lifetime is shown beside each curve. The loss curve of $^{39}$K BECs at 107.75 G in figure (d) is fitted by a double exponential function (orange solid line), in contrast, a single exponential function (black dashed line) cannot fit the data well.}
\label{fig.4}
\end{figure*}
%%%%%%%%%%%%%***FIG. 4***%%%%%%%%%%%%%%%%%%

To better compare the coupling strength and width of the five FRs, we separately measure the loss spectra of single $^{39}$K BECs around each FR. By setting an appropriate holding time of the magnetic field for each FR, the number of residual atoms in the resonance is nearly half of that far from the resonance. As shown in Fig.~\ref{fig.3}, the $g$-wave FR (II) has a very short holding time, indicating that the Feshbach coupling here is stronger, while the resonance (IV) corresponds to a much weaker coupling strength. We use a single Gaussian function to fit the loss spectra in Fig.~\ref{fig.3}, which gives a result of the resonance centers and loss dip widths (defined as the full width at half maximum) of the FRs as summarized in Table~\ref{tab.II}. Note that the width of the loss dip is usually different from the FR width. The loss spectra in Fig.~\ref{fig.3} have a very symmetric line shape, which we think is mainly due to the ultra-low temperature of the atomic gases, the relatively weak coupling strength of the new FRs (for example, compared with the two strong $s$-wave FRs in Fig.~\ref{fig.1}), and the s-wave collision of free atoms in the open channels~\cite{pra.99.022701}.

We also measure the loss curves of atomic gases under different magnetic fields by rapidly ramping the magnetic field strength to the target value and measuring the number of remaining atoms after holding the magnetic field for different times. The experimental results are shown in Fig.~\ref{fig.4}. A single exponential function can fit the experimental data quite well, and the lifetime of the atomic gases is extracted from the fitting. The open circles in Fig.~\ref{fig.4}(a) represent the normal loss process of single $^{39}$K BECs at 117.52 G with the positive scattering length $a=10\ a_{0}$. Atomic gases at this point have the longest lifetime of 8.01 s in Fig.~\ref{fig.4}. The open circles and squares in Fig.~\ref{fig.4}(b, c, d) indicate the fast loss process of $^{39}$K BECs near FRs.

The loss process of the $^{39}$K BECs near FR (III) as shown by the open squares in Fig.~\ref{fig.4}(d) is very interesting. It can be divided into two parts, a very fast loss at the beginning, followed by another greatly slow-downed loss, although there are still many residual $^{39}$K atoms. We fit the loss curve with a double exponential function and extract the atomic lifetimes of the two parts respectively, which gives a result of 0.40 s and 8.01 s. The loss curve of resonance (III) indicates that there is a strong resonance here, but it can only be observed in high atomic density (also see Fig.~\ref{fig.1}(c)) and low temperature (also see Fig.~\ref{fig.1}(b)).

We further examine the effect of mixing $^{87}$Rb atoms on the loss rates of $^{39}$K as shown in Fig.~\ref{fig.4}(a, b, c). Fig.~\ref{fig.4}(a) shows that the loss rate of $^{39}$K atoms is greatly increased due to the presence of a $^{39}$K-$^{87}$Rb interspecies FR at 117.59 G~\cite{pra.77.052705}. Fig.~\ref{fig.4}(b) shows that the loss of $^{39}$K atoms near $g$-wave resonance (II) seems very insensitive to the mixed $^{87}$Rb atoms. In Fig.~\ref{fig.4}(c), the loss rate of $^{39}$K atoms near the d-wave resonance (V) is remarkably decreased, the $^{87}$Rb atoms play the role of buffer gases very well. The different responses of the new FRs to the mixing of $^{87}$Rb atoms may come from the different coupling strengths of FRs and different interspecies and intraspecies scattering lengths in the $^{39}$K and $^{87}$Rb mixture.

In conclusion, we experimentally observed five new FRs of $^{39}$K BECs in $\left|1,-1\right\rangle$ state, two of which are confirmed as $g$-wave and one as $d$-wave FR by the MQDT calculation. So far, only a few HPW FRs have been reported in the literature of $^{39}$K atomic gases. Our findings expand the toolbox for tuning the interaction between $^{39}$K atoms using FRs. Specifically, our newly discovered HPW FRs are located in the region of positive background scattering length, which will be beneficial for studying HPW superfluidity in ultracold $^{39}$K gases~\cite{nature2021.592, pra.97.043626, X2101.05431} . The new HPW FRs have s-wave in open channels and HPW in closed channels, resulting in line shapes of the atomic loss spectra that are completely different from those of FRs with HPW in both open and closed channels in terms of splitting structure and temperature dependence. These FRs have different resonance widths and coupling strengths, and respond differently to atomic temperature or mixed $^{87}$Rb atoms, making it interesting to further study the universal laws of two-body or three-body collision losses around them~\cite{prr.5.013174}.

\begin{acknowledgments}
This research is supported by the National Key Research and Development Program of China (Grants No. 2021YFA1401700, 2022YFA1404101 ), the National Natural Science Foundation of China (Grants No. 12034011, 12374245, 12004229), Innovation Program for Quantum Science and Technology (Grant No. 2021ZD0302003), and the Fund for Shanxi 1331 Project Key Subjects Construction.
\end{acknowledgments}

%%%%%%%%%
%apsrev4-2.bst 2019-01-14 (MD) hand-edited version of apsrev4-1.bst
%Control: key (0)
%Control: author (8) initials jnrlst
%Control: editor formatted (1) identically to author
%Control: production of article title (0) allowed
%Control: page (0) single
%Control: year (1) truncated
%Control: production of eprint (0) enabled

%

%%%%%%%%%

\end{document}